\documentclass{jfm}

\usepackage{commath}
\usepackage[hidelinks]{hyperref}
\usepackage[nameinlink,noabbrev]{cleveref}%
\crefname{figure}{Figure}{Figures}
\crefname{equation}{Equation}{Equations}
\usepackage{graphicx}
\usepackage[numbers]{natbib}
\usepackage[locale=UK,
number-mode=math,
unit-mode=text,
per-mode=symbol,
number-unit-product=\ ,
retain-zero-exponent=false]{siunitx}[=v2]
\DeclareSIUnit{\pixel}{px}
\DeclareSIUnit{\fps}{fps}

\newcommand{\kindex}[2]{\ensuremath{{#1}_{\scalebox{0.5}{#2}}}}

\graphicspath{{../figurematter/latex/figs/}}

\usepackage{todonotes,lipsum}
\usepackage{paralist,palatino,microtype,pifont,ifplatform}
\usepackage{xfrac,amssymb,bm,physics}

\begin{document}
\title{Modelling vortex ring growth in the wake of a translating cone}

\author{Guillaume de Guyon, Karen Mulleners$^{*}$}
\affiliation{Institute of Mechanical Engineering, \'Ecole polytechnique f\'ed\'erale de Lausanne, Switzerland\\
$^*$ Corresponding author: karen.mulleners@epfl.ch
}

\maketitle
\begin{abstract}
Vortex rings have the ability to transport fluid over long distances. They are usually produced by ejecting a volume of fluid through a circular orifice or nozzle. When the volume and velocity of the ejected fluid are known, the vortex' circulation, impulse, and energy can be estimated by the slug flow model. Vortex rings also form in the wake of accelerating axisymmetric bodies. In this configuration, the volume and velocity of the fluid that is injected into the vortex is not known a priori. Here, we present two models to predict the growth of the vortex behind disks or cones. The first model uses conformal mapping and assumes that all vorticity generated ends up in the vortex. The vortex circulation is determined by imposing the Kutta condition at the tip of the disk. The position of the vortex is integrated from an approximation of its velocity, given by Fraenkel. The model predicts well the maximum circulation of the vortex, but does not predict the tail shedding observed experimentally. A second model is based on an axisymmetric version of the discrete vortex method. The shear layer formed at the tip of the cone is discretised by point vortices, which roll-up into a coherent vortex ring. The model accurately captures the temporal evolution of the circulation and the non-dimensional energy. It also predicts the occurrence of tail shedding and the total amount of vorticity lost in the wake. The portion of the lost vorticity due to tail shedding is sensitive to the choice of the numerical parameters controlling the stability of the shear layer.
\end{abstract}

\section{Introduction}
Vortex rings can be created by ejecting a volume of fluid through a circular orifice.
A shear layer forms at the exit of the orifice, rolls-up, and forms a coherent axisymmetric ring shaped vortex.
Mesmerising examples of such vortex rings can be found in nature.
Some aquatic animals create vortex rings in their wake to enhance their propulsion \citep{Johnson1972,Linden2004}, and dolphins even produce them for entertainment \citep{Marten1996}.
Smoke vortex rings that are generated by smoke erupting from volcanoes can travel at more than a hundred meters per seconds and climb up to hundreds of meters without loosing their coherence \citep{Suwa2014}.
The general stability and efficiency of vortex rings at transporting fluid has inspired engineers.
Chimneys are adapted to intermittently release smoke to produce vortex rings to increase the rise altitude of pollutants \citep{Turner1960}.
Cavitating vortex rings are used to perform underwater cleaning or to cut rocks \citep{Chahine1983}.

The classical apparatus to study the formation of vortex rings is a piston-cylinder.
A piston ejects fluid through a circular nozzle and generates vortex rings in a repeatable and controlled way.
The temporal evolution of the vortex rings is typically described in terms of the non-dimensional time $T^*=L/D$ which corresponds to the ratio between the travelled piston stroke length $L$ and the nozzle diameter $D$.
The growth and shedding of a vortex ring generated by a piston-cylinder apparatus can be predicted by the so called slug flow model \citep{Gharib1998b}.
Based on the assumption that all the fluid ejected by the piston ends up in the vortex, the circulation $\Gamma$, impulse $I$, and energy $E$ of the vortex can be expressed as a function of the stroke length and piston velocity.
For example, the non-dimensional energy, defined as $E^*=E/\sqrt{I \Gamma^3}$, is given by $E^*=\sqrt{\pi/2}/ T^*$ \citep{Gharib1998b}.
For a broad range of piston translation profiles, the vortex rings reach a critical minimum value of $E^*=0.31$ around $T^* \approx 4$ when the vortex ring separates from its feeding shear layer \citep{Gharib1998}.
This non-dimensional vortex separation time is called the vortex formation number.
The prediction of the exact vortex formation number by the slug flow model has recently been improved to account for different orifice geometries, by including the effect of the flow contraction at the exit of the nozzle~\citep{Limbourg2021}.

The translational velocity of a vortex ring depends on its non-dimensional energy.
For a viscous steady vortex ring, the velocity $U_0$ of a vortex ring with diameter $D_0$ can be estimated according to \citet{Saffman1970} as:
\begin{equation}
U_0 = \dfrac{\Gamma}{\pi D_0}\left( E^* \sqrt{\pi} +\dfrac{3}{4} \right).
\label{eq:uo}
\end{equation}
The experimentally observed range of vortex formation times corresponding to $0.27<E^*<0.4$ would translate to vortex ring velocities between \SI{50}{\percent} and \SI{60}{\percent} of the piston velocity when considering the \citeauthor{Norbury1973} family of vortices \citep{Norbury1973, Gharib1998b}.
Approximations of the vortex velocity using \citeauthor{Fraenkel1970}'s second order expansions lead to similar results \citep{Fraenkel1970, Shusser2000}.
This range of vortex velocities is close to the shear layer velocity, which is assumed in first approximation to be half of the piston velocity.
From a kinematic point of view, \citeauthor{Gharib1998b} concluded that the vortex ring separates when its translational velocity exceeds the velocity of its feeding shear layer.

Vortex rings can also emerge in the wake of accelerating bodies, such as discs \citep{Rosi2017} or cones \citep{DeGuyon2021}.
In this configuration, the vortex self induced velocity is directed towards the vortex generator, and the vortex does not move away from its feeding shear layer.
Separation occurs at larger time scales, when azimuthal instabilities break the symmetry of the vortex \citep{Johari2002}.
Yet, the non-dimensional circulation and non-dimensional energy of these drag vortex rings converge within the same convective times scales as their propulsive counterparts \citep{DeGuyon2021}.
To predict the growth of drag vortices, we cannot revert to the slug flow model as the slug of fluid that enters the vortex is not known a priori.
A classical alternative approach to predict vortex growth consists in simulating directly the roll-up of an inviscid vortex sheet representing the shear layer \citep{DeVoria:2018cy}.
Early simulations of the growth of a vortex behind an impulsively starting flat plate showed that the circulation grows in time as $t^{1/3}$ \citep{Pullin1978}.
The discretisation of the vortex sheet by point vortices accurately reproduces the vortex shedding in the wake of objects, such as flat plates or airfoils experiencing massive flow separation \citep{Katz1981a,Xia2013,Darakananda:2019ks}.

Most studies focus on two-dimensional flows, and the few models developed for three-dimensional axisymmetric vortex rings were primarily used to simulate the piston-cylinder experiment \citep{Acton1980,Nitsche1994}.
The roll-up of axisymmetric shear layers display some interesting features that are absent in the two-dimensional configuration.
A continuously evolving ring vortex will release or loose part of its vorticity in its wake.
This phenomenon is known as tail shedding and is more pronounced in an axisymmetric than in a two-dimensional configuration \citep{Krasny2002a,Ofarrell2012}.
To accurately describe the growth of an axisymmetric vortex ring, we should estimate the onset and the role of tail shedding.

Here, we develop a low order model of the vortex growth behind translating disks and cones.
The vortex in the wake of the disk is first modelled by applying conformal mapping to a circular vortex filament.
The vortex circulation is found by enforcing the Kutta condition at the edge of the disk, and a correction is applied to the circulation to account for the fact that conformal mapping does not preserve irrotationality in the axisymmetric plane.
Second, we develop a discrete vortex method for axisymmetric flows.
The growth of a vortex in the wake of cones is computed and compared with experimental results presented in \cite{DeGuyon2021}.
The amount of vorticity lost due to tail-shedding and its effect on the vortex growth are quantified for various numerical parameters.

\section{Experimental setup}
\begin{figure}
\includegraphics{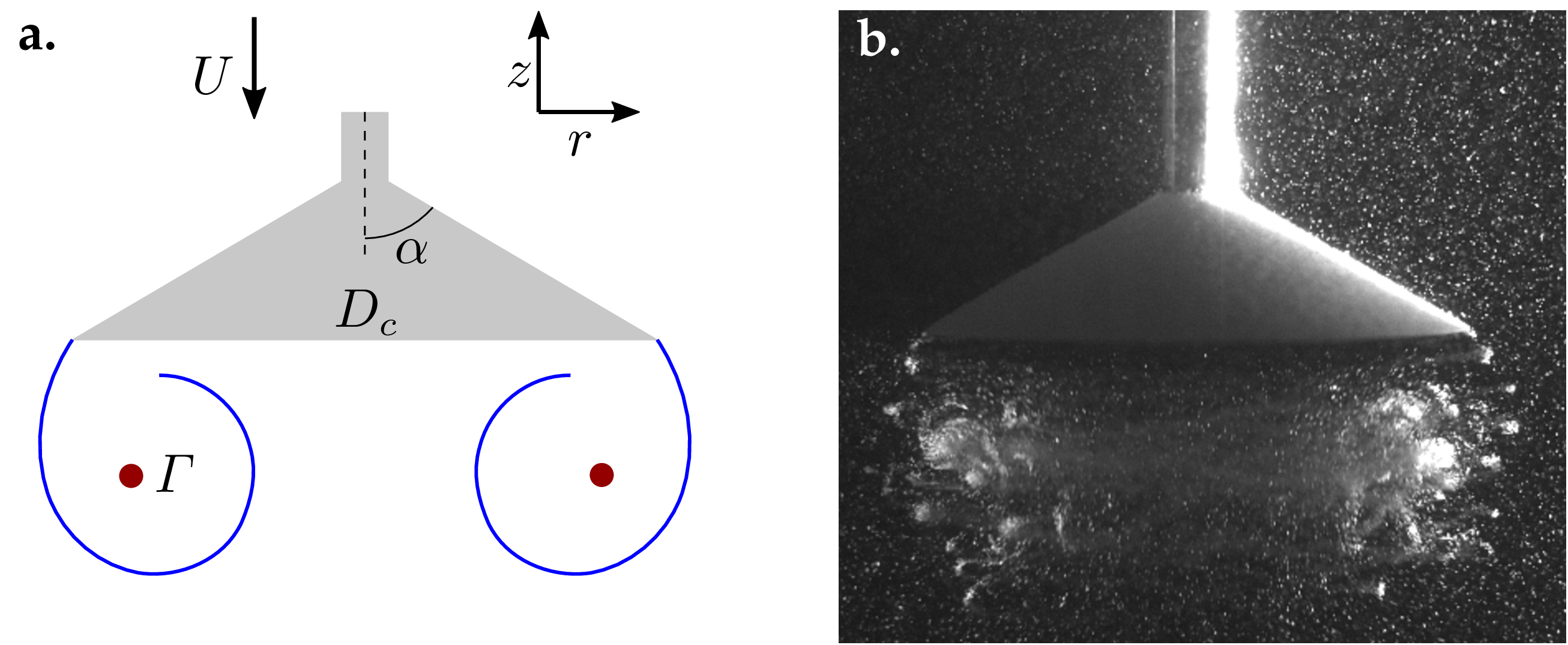}
\caption{\textbf{a}, Schematic of the cone and vortex parametrisation. \textbf{b}, Raw camera image of the roll up of the shear layer into a vortex ring, visualised by seeding particles.}
\label{fig:setup}
\end{figure}
A cone immersed in water is pulled upward in quiescent fluid along its symmetry axis (\cref{fig:setup}a).
The translation is performed by a NEMA 17 stepper motor connected to a belt driven linear actuator.
The cones are 3D printed, their diameter $D_c$ ranges from \SIrange{3}{9}{cm} and the aperture $\alpha$ from \SIrange{30}{90}{\degree}.
The cones are accelerated at \SI{3}{\meter\per\second\squared} from rest, up to velocities $U$ ranging from \SIrange{0.35}{0.7}{\meter\per\second}.
A total of twenty experiments were carried out, at diameter based Reynolds numbers $Re^{}_D=UD_c/\nu$ between \num{1e4} and \num{6e4}.

The flow in a symmetry plane of the cone is illuminated with two light emitting diodes (LED) and the in-plane components of the velocity fields are measured using time-resolved particle image velocimetry (PIV).
A total field of view of \SI{18x36}{\centi\meter} is recorded at \SI{1000}{fps} by two high speed cameras, each with a resolution of \SI{1024x1024}{px}, placed on top of each other.
The images are processed with a multi-grid algorithm and a final interrogation window size of \SI{24x24}{px} with an overlap of \SI{60}{\percent}, leading to a physical resolution of \SI{1.8}{\milli\meter}, or \SI{6}{\percent} of the smallest cone diameter.

The non-dimensional time of the experiment is defined by the distance $L$ travelled by the cone relative to its diameter: $T^*=L/D_c$.
Particle image velocimetry gives access to the velocity field $(u,v)$, from which the out-of-plane component of the vorticity field $\omega$ is derived.

\begin{figure}
\centering
\includegraphics{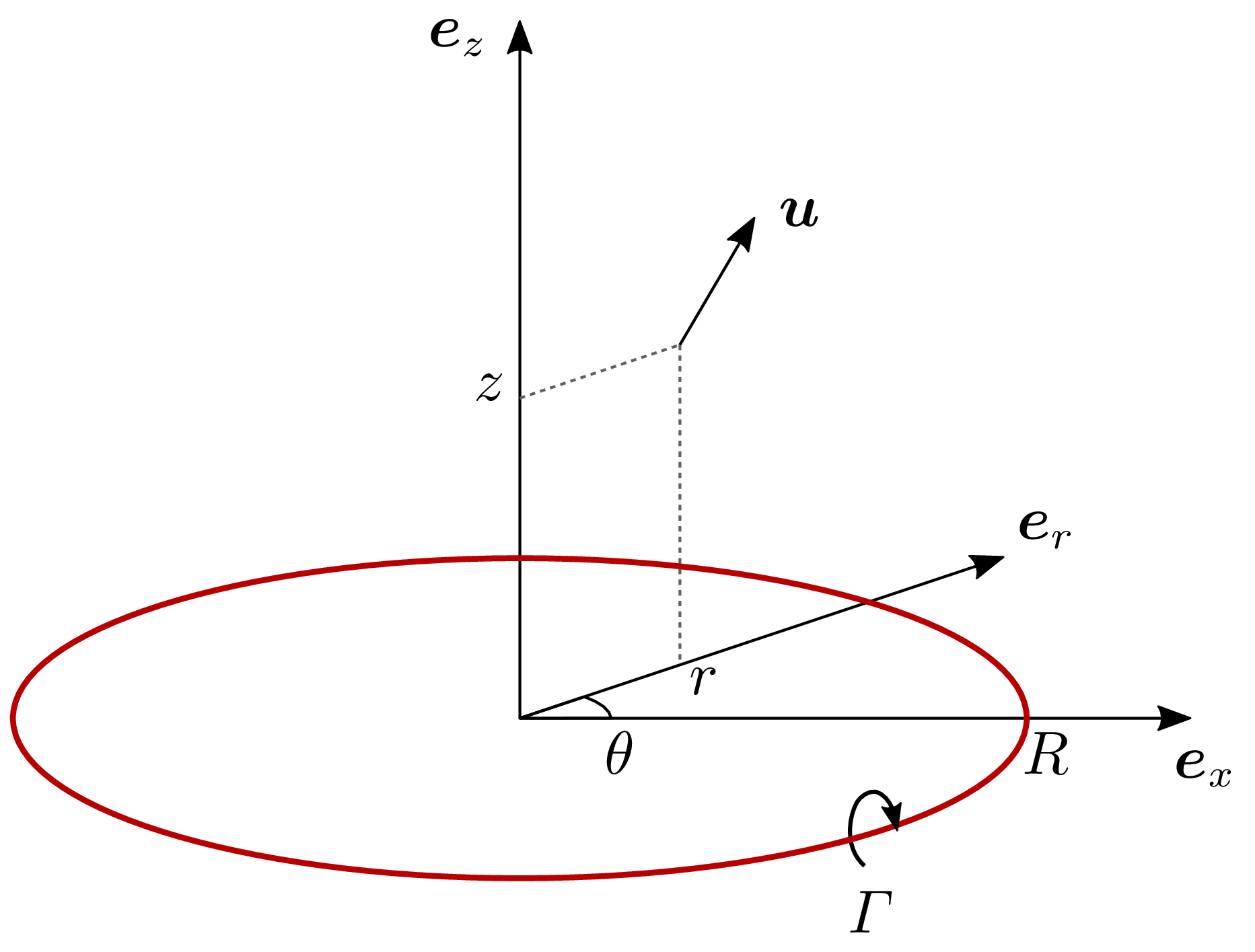}
\caption{Velocity $\bm{u}$ induced by an infinitely thin vortex ring of circulation $\Gamma$ and radius $R$ on a point of coordinates $(r,z)$.}
\label{fig:velind}
\end{figure}

\section{Numerical methods}
\subsection{Circular vortex filaments}
Any incompressible flow field $\bm{u}$ can be derived from a vector potential $\bm{A}$.
The velocity field associated with a vortex ring is axisymmetric in the cylindrical coordinate system $(\bm{e}_r,\bm{e}_{\theta},\bm{e}_z)$, with a zero azimuthal velocity component.
As a consequence, the vector potential reduces to the scalar stream function $\psi$:
\begin{equation}
	\bm{u}=\nabla \times \bm{A} \quad \text{ with } \quad \bm{A}=(0,A,0)=\left(0,\dfrac{\psi}{r},0\right) \quad .
\end{equation}
The vorticity is reduced to its azimuthal component,
\begin{equation}
	\bm{\omega} = \nabla \times \left( \nabla \times \bm{A} \right)=\left(0,\omega,0\right)
	\quad \text{ with } \quad
	\omega = -\pdv[2]{A}{r}-\pdv[2]{A}{z}-\pdv{A/r}{r} \quad .
	\label{eq:omeg}
\end{equation}
In analogy to the point-vortex model in two dimensions, we can approximate a three-dimensional vortex ring by a circular vortex filament (\cref{fig:velind}).
In the $(\bm{e}_r,\bm{e}_z)$ plane, the vorticity field of a circular ring described by $\mathbf{r}_v=(r_v,z_v)$ and circulation~$\Gamma$ is given by $\omega(\mathbf{r})=\Gamma \delta(\mathbf{r}-\mathbf{r}_v)$, with $\delta$ the Dirac function.
This vorticity distribution leads to an analytical solution to \cref{eq:omeg}, in an unbounded fluid.
The velocity and stream function induced by this vortex are expressed as a function of the kernels $\bm{k}^u$ and $k^{\psi}$ \citep{Yoon2004} such that:
\begin{equation}
	\bm{u}(\mathbf{r},\mathbf{r}_v,\Gamma)=\Gamma \bm{k}^u(\mathbf{r},\mathbf{r}_v) \quad \text{and} \quad \psi(\mathbf{r},\mathbf{r}_v,\Gamma)=\Gamma k^{\psi}(\mathbf{r},\mathbf{r}_v)\quad.
\end{equation}
The kernels are given by:
\begin{align}
\label{eq:d1}
	\bm{k}^u&=
	\begin{bmatrix}
		k^u_r\\
		k^u_z
	\end{bmatrix}
	=\dfrac{r_v}{\pi (1-m)a^3}
	\begin{bmatrix}
		(z-z_v)D\\
		r_v E-rD
	\end{bmatrix}\\
	k^{\psi}&=\dfrac{a}{4\pi} \left(\left(2-m\right)K-2E\right)
\label{eq:d2}
\end{align}
with $a^2=(r_v+r)^2+(z-z_v)^2$, $m=4rr_v/a^2$ the elliptical modulus, $K(m)$ and $E(m)$ the complete elliptic integral of the first and second kind, and ${D=2(K-E)\left(1-m^{-1}\right)+E}$.

\begin{figure}
	\centering
	\includegraphics{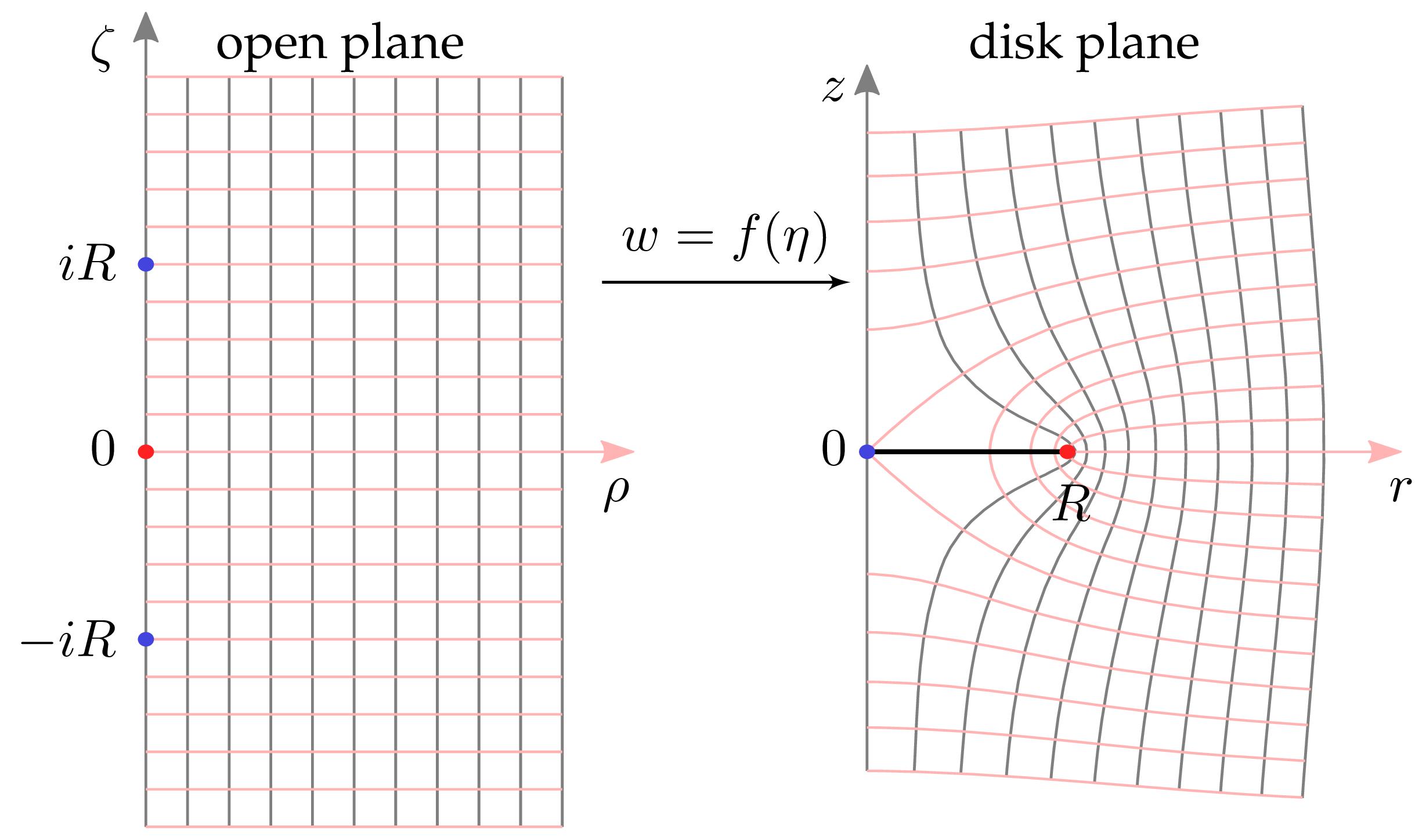}
	\caption{Mapping of the half plane $\eta=\rho +i \zeta$ with $\rho>0$ to the half plane $w=r+iz$ by the function $f(\eta)=\sqrt{\eta^2+R^2}$}
	\label{fig:mapmap}
\end{figure}
\subsection{Conformal mapping of the disk}\label{sec:CM}
The flow described by \cref{eq:d1,eq:d2} is valid in an unbounded fluid.
To account for the presence of a disk, we present here a solutions using conformal mapping.
The function $f(\eta)=\sqrt{\eta^2+R^2}$ maps the half plane $\eta=\rho+i \zeta$ into the half plane $w=r+iz$ with a disk of radius $R$ (\cref{fig:mapmap}).
The inverse mapping function is $g(w)=\sqrt{z^2-R^2}$.
The subscripts $o$ and $d$ refer to a quantity expressed respectively in the  open plane $\eta$ and in the disk plane $w$.

The use of conformal mapping in cylindrical coordinates requires some precaution.
In Cartesian coordinates $(\bm{e}_x,\bm{e}_y,\bm{e}_z)$, a two-dimensional incompressible flow in the $(\bm{e}_x,\bm{e}_z)$ plane is derived from a potential vector $\bm{\psi}=(0,\psi,0)$, with 	$\bm{u}=\nabla \times \bm{\psi}$.
The vorticity is $\bm{\omega}=(0,\omega,0)=\nabla \times \left(\nabla \times \bm{\psi}\right)$, which is equivalent to ${\omega = -\kindex{\Delta}{2D} \psi}$.
The two-dimensional Laplacian operator $\kindex{\Delta}{2D}$ is defined as $\pdv*[2]{x}+\pdv*[2]{z}$ or $\pdv*[2]{r}+\pdv*[2]{z}$, depending on the coordinate system.
We now consider the conformal mapping $g(w) = \eta$ and note $\psi_o$ and $\psi_d$ the stream functions in the open and disk planes, with $\psi_o(w)=\psi_d\left(g(w)\right)$.
For simplicity, we merge the notation of the functions on their complex and Cartesian supports: $\psi_d(r+iz)=\psi_d(r,z)$.
The Cauchy-Riemann relation leads to
\begin{equation}
\kindex{\Delta}{2D} \psi_d = |g'(w)|^2 \kindex{\Delta}{2D} \psi_o \quad.
\end{equation}
Consequently, the irrotationality of the flow is preserved and a point vortex in one plane maps onto a point vortex in the other plane.

The situation is different in cylindrical coordinates where \cref{eq:omeg} boils down to
\begin{equation}
	\omega = -\kindex{\Delta}{2D} A -\pdv{A/r}{r} .
	\label{eq:omeg2}
\end{equation}
To ensure the incompressibility of the flow and the non-penetration condition, we impose that the potential vector is preserved by the mapping: $A_d(w)=A_o(g(w))$.
From the Cauchy-Riemann relations, we now obtain:
\begin{equation}
\kindex{\Delta}{2D} A_d = |g'(w)|^2 \kindex{\Delta}{2D} A_o \quad,
\end{equation}
and \cref{eq:omeg2} leads to
\begin{equation}
	\omega^{}_d+ \pdv{A_d/r}{r} = |g'(w)|^2 \left(\omega^{}_o+\pdv{A_o/\rho}{\rho}\right)
\end{equation}
with $\omega_d$ and $\omega_o$ the vorticity in the open and disk planes.
The open plane contains only a uniform flow and a point vortex and the flow is irrotational ($\omega_o=0$ everywhere).
The situation is different in the disk plane where the vorticity is:
\begin{equation}
	\omega_d = |g'(w)|^2 \pdv{A_o/\rho}{\rho}- \pdv{A_d/r}{r} \neq 0 \quad .
	\label{eq:IG}
\end{equation}

\begin{figure} 
	\centering
	\includegraphics{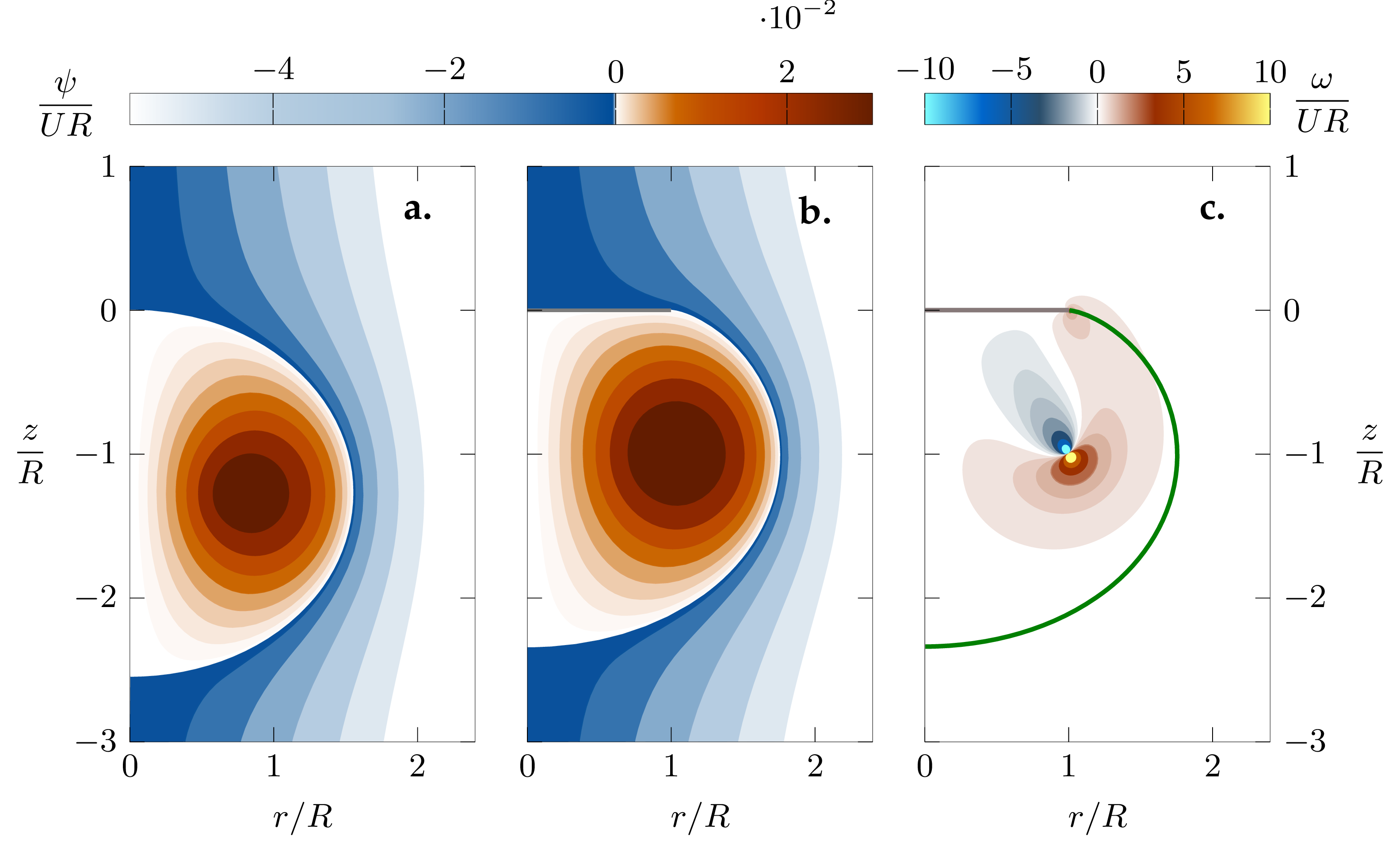}
\caption{Contours of the stream function in \textbf{a}, the open plane and in
\textbf{b}, the disk plane.
\textbf{c}, Vorticity created by the axisymmetric conformal mapping between the open and the disk plane.
The green line indicates the $\psi=0$ contour.}
	\label{fig:map}
\end{figure}

The last step in the modelling of the vortex in the wake of the disk is to ensure that the Kutta condition is met at the extremity of the disk.
The condition $u_d(R,0)=0$ in the disk plane is equivalent to $u_o(0,0)=0$ in the open plane and \cref{eq:d1,eq:d2} lead to
\begin{equation}
	\Gamma_o = 2U\dfrac{|\eta|^3}{\rho^2}
	\label{eq:Geta}
\end{equation}
with $\eta=\rho+i\zeta$ the position of the point vortex in the open plane.
An example of the conformal mapping for a vortex located at $(R,-R)$ behind the disk is presented in \cref{fig:map}.
The contours of the stream functions in the open and disk planes are presented in \cref{fig:map}a-b.
The vorticity created by the mapping (\cref{fig:map}c) is mainly concentrated near the vortex centre and near the extremity of the disk.
The circulation $\Gamma_d$ obtained by integrated \cref{eq:IG} is $\SI{10}{\percent}$ lower than the original circulation $\Gamma_o$.
This decrease depends on the vortex position and its influence on the results is further discussed in the results (\cref{sec:lom}).

\subsection{Discrete vortex method}\label{sec:DVM}
\begin{figure}
	\centering
	\includegraphics[scale=0.8]{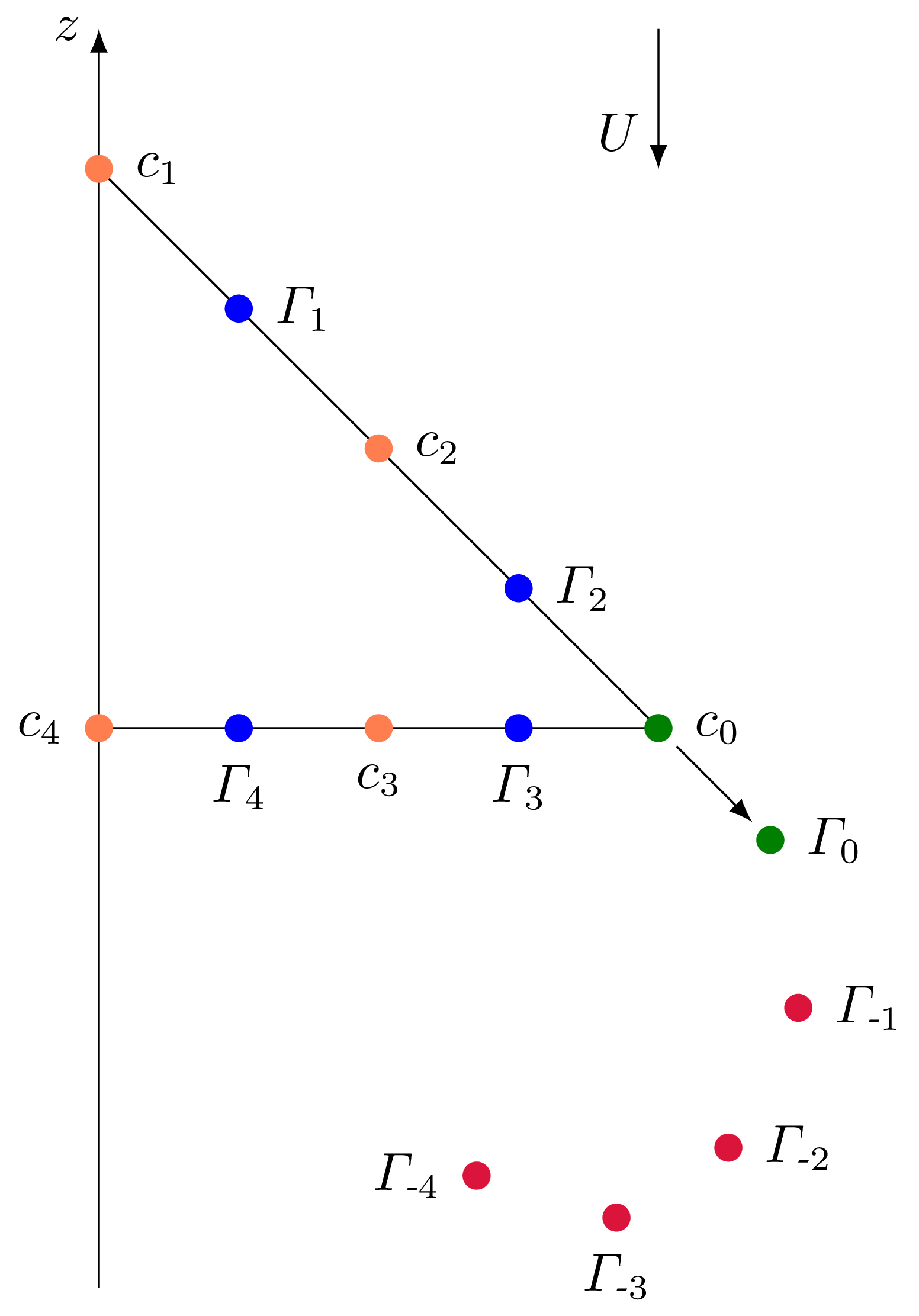}
	\caption{Schematic representation of the discrete vortex model.
The cone itself is discretised using blue vortices $\Gamma_1 ... \Gamma_4$.
The vortices $\Gamma_{0} ... \Gamma_{\text{-}4}$ in red discretise the shear layer forming at the tip of the cone.
On control points $c_1 ... c_4$, the velocity is tangent to the cone's surface.}
\label{fig:model3}
\end{figure}

To complement the conformal mapping based approach, we develop a discrete vortex method for the axisymmetric flow to study the loss of vorticity due to tail shedding.
The cone is discretised by $N$ point vortices of unknown circulations $\Gamma_1...\Gamma_N$ (\cref{fig:model3}).
A number of $N$ control points $c_1...c_N$ are set in between these vortices and on the $z$ axis.
On a control point $c_i$, the unit vector normal to the cone is noted by $\mathbf{n}_i$.
The non-penetration condition on the cone imposes that the normal velocity is zero at each control point.
The stagnation streamline at the tip of the cone at the control point $c_0$ is set to align with the generatrix of the cone.
This condition is sometimes referred as the Giesing-Maskell model \citep{Giesing1969,Maskell1971,Xia2017}.
The normal to the generatrix is noted by $\mathbf{n}_0$ and the condition is enforced by releasing a vortex $\Gamma_0$ at a specific location.
The vortices released at previous time steps are noted with negative indices, $\Gamma_{\text{-}1} ... \Gamma_{\text{-}p}$.
The non-penetration and tip conditions lead to a linear system of $N+1$ equations of unknowns $\Gamma^{}_0 ... \Gamma^{}_N$:
\begin{equation}
	\forall j \in \{0,\ldots,N\}, \quad \bm{u}.\mathbf{n}_j+\sum_{i=-p}^N \Gamma_i \bm{k}_{ij}^u.\mathbf{n}_j= 0
	\label{eq:lin1}
\end{equation}
with $\bm{k}_{ij}^u$ the Biot-Savart kernel (\cref{eq:d1,eq:d2}) from a vortex $\Gamma_i$ to a control point $c_j$.

At each time step, the circulation of the panel vortices are recomputed, and the flow vortices are convected with a fourth-order Runge-Kutta method.
For the convective part, a desingularisation is applied to the velocity kernel for the flow vortices, by adding a smoothing term $\epsilon$ to \cref{eq:d1,eq:d2}:
\begin{equation}
\label{eq:smoothparam}
a^2=(r_v+r)^2+(z-z_v)^2+\epsilon^2\quad.
\end{equation}

\section{Results}
\begin{figure}
\includegraphics{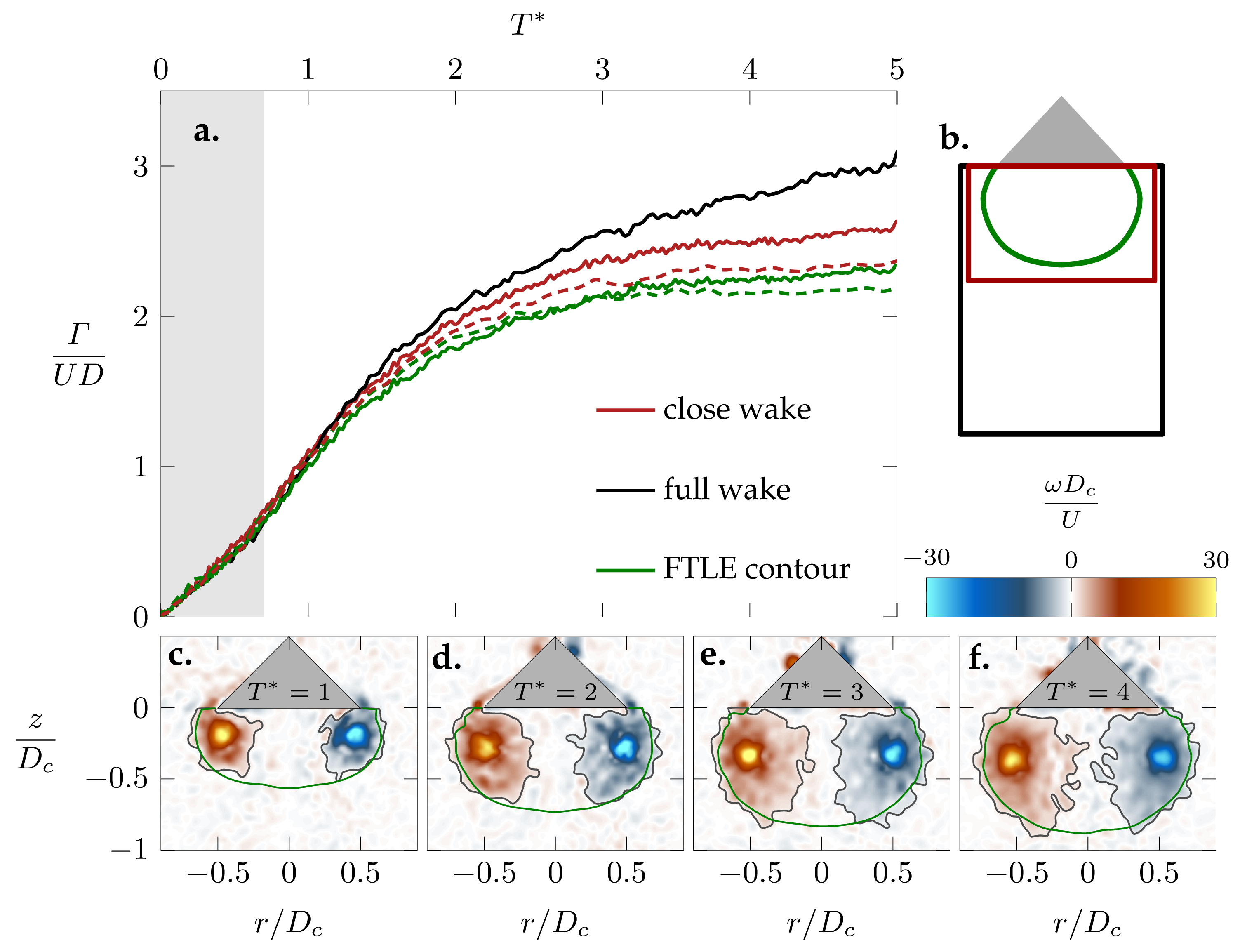}
\caption{\textbf{a}, Temporal evolution of the circulation of the vortex and in the wake of the cone.
The grey area indicates the acceleration phase of the cone.
The circulation is non-dimensionalised by the cone diameter (solid lines) or the vortex diameter (dashed lines).
The line colours indicate the integration area that is considered for the calculation of the circulation.
\textbf{b}, Contours bounding the different integrations areas, including the FTLE contour enclosing the vortex, a small rectangular contour directly downstream of the cone, and a larger rectangular contour capturing the full wake.
\textbf{c}-\textbf{f}, Instantaneous snapshots of the vorticity field including the FTLE contour (green line) and the vorticity isocontour $\omega=U/D_c$ (grey line), in the wake of a translating cone with $\alpha=\ang{45}$, $D_c =\SI{6}{\centi\meter}$ and $U = \SI{0.5}{\meter\per\second}$.}
	\label{fig:circ1}
\end{figure}

\subsection{Experimental results}
Exemplary snapshots of the vortex development for a translating cone with $\alpha=\ang{45}$, $D_c =\SI{6}{\centi\meter}$ and $U = \SI{0.5}{\meter\per\second}$ are presented in \cref{fig:circ1}c-f.
The vortex contour is delimited by a ridge of the positive finite time Lyapunov exponent (pFTLE).
This Lagrangian method outlines the material boundaries of the flow field and has been used to detect vortices from various vortex generators \citep{Green2011,Shadden2006,OFarrell2010,Krishna2018}.
The position of the vortex centre $(R_v,Z_v)$ in the cylindrical coordinate system is calculated as
\begin{align}
\label{eq:centerr}
R_v^2 & =\frac{\iint \omega r^2 \dd{r} \dd{z}}{\iint \omega \dd{r} \dd{z}} ,\qquad D_v=2R_v\\
Z_v & =\frac{\iint \omega zr^2 \dd{r} \dd{z}}{\iint \omega r^2 \dd{r} \dd{z}} \quad.
\label{eq:centerz}
\end{align}
The positions are given relative to the base of the cone.

Vorticity is generated on the conical object and is fed into the vortex through the shear layer.
Some of the vorticity produced at the tip of the cone circumvents the vortex, which is bounded by the FTLE contour.
This is outlined by the iso-contour of vorticity $\omega =U/D_c$  on \cref{fig:circ1}(c-f), which extends outside of the FTLE contour.

The amount of vorticity circumventing the recirculation zone associated with the vortex is quantified by computing the circulation
\begin{equation}
\Gamma=\iint \omega \dd{r} \dd{z}
\label{eq:circ}
\end{equation}
on different contours (\cref{fig:circ1}a,b):
\begin{inparaenum}
\item the recirculation area delimited by the FTLE ridge,
\item a rectangular contour of width $1.8\,D_c$ and height $D_c$ in the near wake of the cone, and
\item a larger rectangular contour capturing the full wake behind the cone.
\end{inparaenum}
The circulation is non-dimensionalised by the cone diameter (solid lines) or the vortex diameter (dashed lines).

The circulation in the wake of the cone immediately starts to increase when the cone is accelerated from rest.
Until $T^*=1$, all circulation ends up in the vortex within the FTLE contour.
The initial rate of increase of the circulation is approximately constant ($\dot{\Gamma}=1.4U^2$) and decreases in time.
Between $T^*=1$ and $T^*=3$, the rates of increase in circulation inside the different contours diverge and drop to new constant values for $T^*>3$ of $\dot{\Gamma}=0.06U^2$ within the FTLE contour and $\dot{\Gamma}=0.2U^2$ in the entire wake.
When the circulation is non-dimensionalised with the vortex diameter, instead of the cone diameter, the vortex circulation converges to a constant value of $\Gamma/(UD_v)\approx2.3$ after three convective times.
This convergence has been observed for all experiments with different cone geometries and translation velocities \citep{DeGuyon2021}.
The vortex circulation converges when vorticity inside the vortex has spread radially towards the axis of symmetry and there is no more non-vortical fluid within the FTLE contour.
The divergence between the growth rate of the circulation inside the vortex and in the full wake after $T^*>1$ indicates that some of the vorticity produced is no longer entering the vortex or some of it is leaving the vortex again.
The circulation in the near wake contour is around $\SI{10}{\percent}$ higher than the circulation enclosed by the FTLE contour, but the rate of increase in both near wake contours reaches the same value after three convective times.
The excess circulation in the full wake is thus due to vorticity that is convected downstream and is not the result of a radial spread of vorticity outside the vortex area bound by the FTLE contour.

\begin{figure}
	\includegraphics{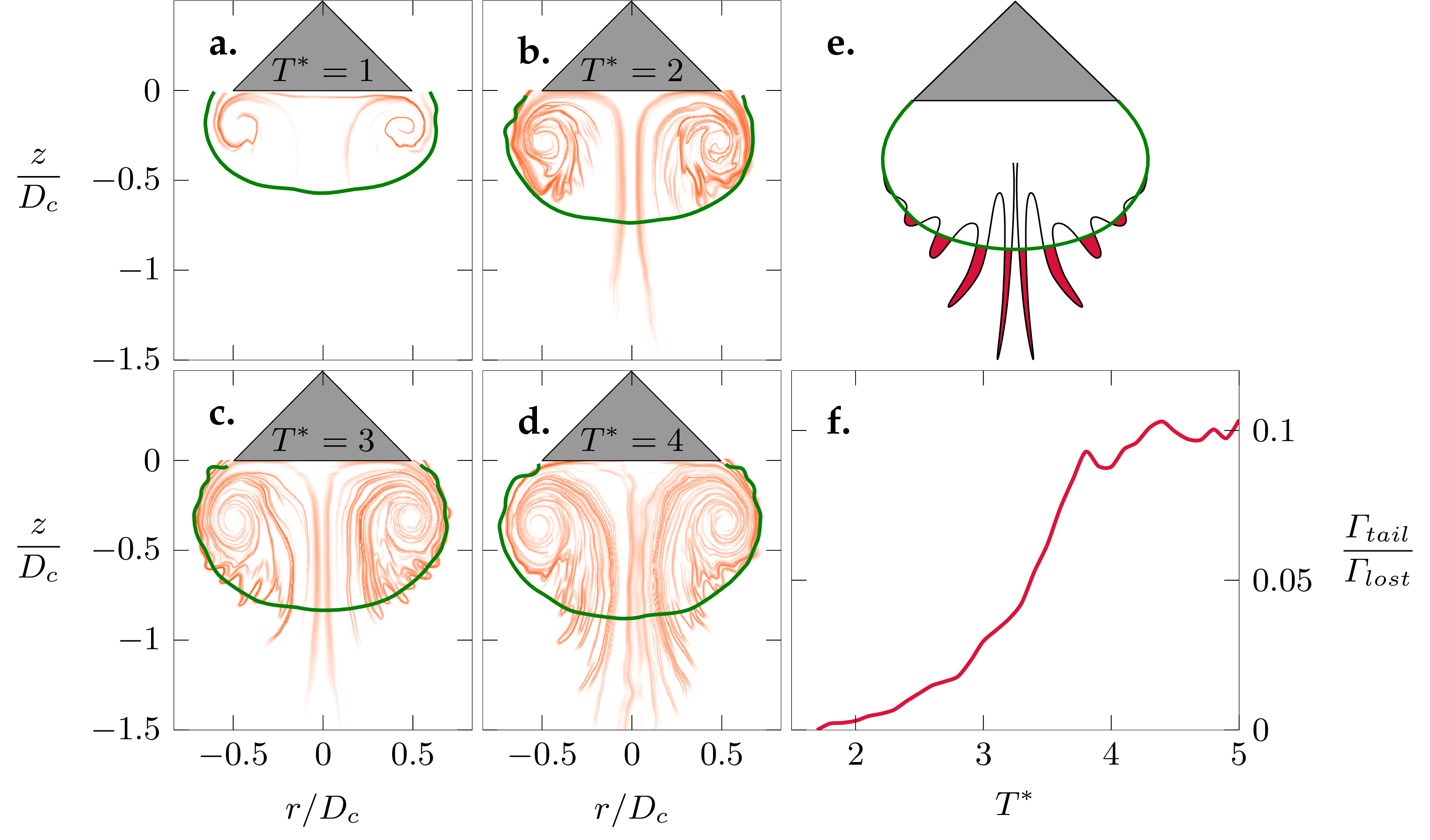}
	\caption{\textbf{a-d}, Snapshots of the ridges of the FTLE fields. In red, the backward nFTLE ridges, in green, the forward pFTLE ridge.
\textbf{e}, Schematic of the lobes of the nFTLE that protrude underneath the pFTLE boundary.
The circulation in the shaded areas is identified as the tail circulation.
\textbf{f}, Temporal evolution of the tail circulation, relative to the lost circulation defined as the difference between the total circulation in the wake and the circulation in the vortex.}
	\label{fig:ftle_n_p}
\end{figure}
We define the lost circulation $\kindex{\Gamma}{lost}$ as the difference between the total circulation in the wake of the cone and the circulation inside the forward or positive time pFTLE contour.
The vorticity lost in the wake can have two origins.
Either the fluid particles directly circumvent the recirculation area of the vortex, or fluid particles from within the vortex are ejected after a few rotations.
The latter fluid detrainment mechanism is known as tail shedding \citep{Shariff2006}.
It can be visualised by combining the ridges of the backward and forward FTLE field (\cref{fig:ftle_n_p}).
The lobes of the backward or negative time nFTLE ridges are progressively growing outside the vortex boundary marked by the pFTLE ridge \cref{fig:ftle_n_p}c-d.
The volume of fluid and the vorticity they carry, is ejected from the vortex (\cref{fig:ftle_n_p}e).
The temporal evolution of the vorticity lost via tail shedding, or tail circulation $\kindex{\Gamma}{tail}$, with respect to the total lost circulation, $\kindex{\Gamma}{lost}=\kindex{\Gamma}{wake}-\kindex{\Gamma}{vortex}$ is presented in  \cref{fig:ftle_n_p}f.
The tail circulation starts to increase shortly before $T^*=2$ and increases substantially after $T^* \approx 3$ and converges to steady state value of \SI{10}{\percent} of the total lost circulation.
The tail circulation is measured by integrating the vorticity within the nFTLE lobes protruding outside the vortex boundary.
The accuracy of this approach depends on the quality of the lobe detection.
By adjusting the lobe detection to be more restrictive or more inclusive, we estimate the uncertainty and find that \SIrange{5}{15}{\percent} of the lost circulation is due to tail shedding.
This result is in line with the findings of the numerical study by \cite{OFarrell2014}, who showed that a vortex ring in an unbound flow can lose up to \SI{14}{\percent} of its circulation via tail shedding.

\subsection{Low order model of the vortex growth rate}\label{sec:lom}
\begin{figure}
\includegraphics{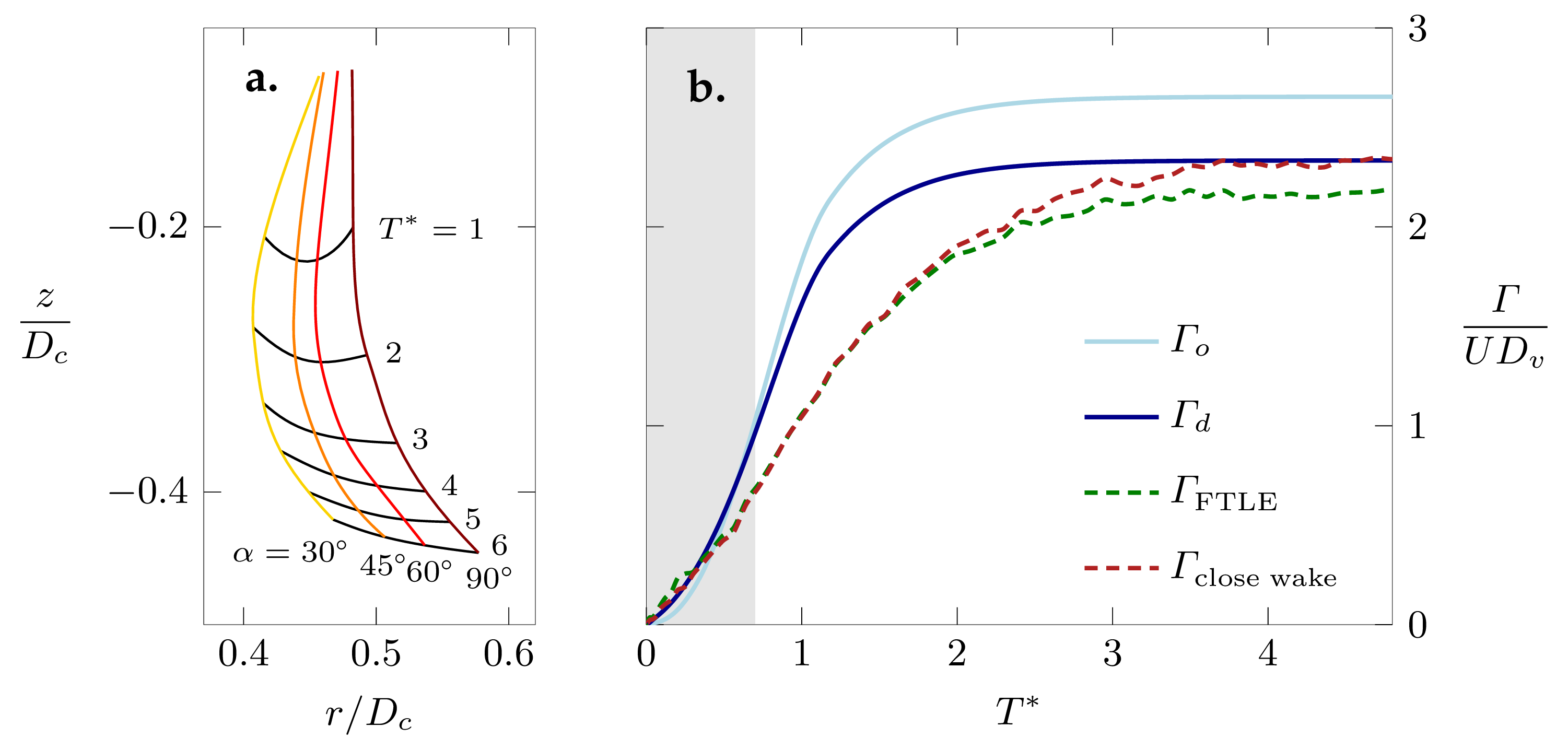}
\caption{\textbf{a}, Trajectories of the vortex centre for different cone apertures.
\textbf{b}, Temporal evolution of the circulation in the disk plane and the open plane compared to the experimentally measured circulation.}
\label{fig:fitE}
\end{figure}

With the help of the conformal mapping tools developed in \cref{sec:CM}, we model the temporal evolution of the vortex in the wake of a starting disk by computing its circulation and position $w$.
The model subsequently goes through the following steps.
First, the vortex location $w$ in the disk plane is mapped to its location $g(w)$ in the open plane.
Second, the vortex circulation is determined by applying the Kutta condition yielding
\begin{equation}
\Gamma_o = 2U(t)\dfrac{|g(w)|^3}{\operatorname{Re}(g(w))^2}
\end{equation}
where $\operatorname{Re}$ denotes the real part.
The circulation in the open plane is then converted to the circulation in the disk plane by integrating \cref{eq:IG}.
Finally, the trajectory of the vortex is determined by integrating
\begin{equation}
\dfrac{dz_v}{dt}=U_v
\label{eq:dzv}
\end{equation}
with $U_v$ the self-induced velocity of the the vortex.
The velocity $U_v$ is expressed using \citeauthor{Fraenkel1970}'s second order approximation for vortices of small cross-section:
\begin{equation}
U_v=\dfrac{\Gamma}{4\pi R_v} \mathcal{F}(\varepsilon) \quad \text{with} \quad
\mathcal{F}(\varepsilon) = \log{\dfrac{8}{\varepsilon}}-\dfrac{1}{4}+\left(\dfrac{\varepsilon}{8}\right)^2 \left(37\log{\dfrac{8}{\varepsilon}}-\dfrac{223}{4} \right).
\label{eq:Uth}
\end{equation}
In \citeauthor{Fraenkel1970}'s approximation, the parameter $\varepsilon$ is the vortex core radius relative to the ring radius. It corresponds to $z_v/R_v$ in the present experiment.
Here, the main vortex motion is parallel to the cone's motion.
The radial excursions of the vortex centre position are lower in magnitude.
The vortex radius remains close to the disk radius for various cone apertures (\cref{fig:fitE}a) and we assume $R_v=R$ for the remainder of the section.
\cref{eq:dzv,eq:Uth} still have two unknowns, the stream-wise location of the vortex $z_v$ and its circulation $\Gamma$:
\begin{equation}
	\dfrac{dz_v}{dt}=\dfrac{\Gamma}{4\pi R}\mathcal{F}(\varepsilon) \quad \text{with}\quad \varepsilon=\dfrac{z_v}{R} \quad.
		\label{eq:Uth2}
\end{equation}
We use the conformal mapping tools developed in \cref{sec:CM} further to estimate the circulation.
The position of the vortex $w=R-iz_v$ in the disk plane maps to $g(w)=\sqrt{z_v^2-R^2}$ in the open plane.
The circulation is set to fulfil the Kutta condition at the tip of the disk, $u_z(R,0)=0$, which, based on \cref{eq:Geta}, is equivalent to imposing
\begin{equation}
	\Gamma_o=2U(t)\dfrac{|g(w)|^3}{\operatorname{Re}(g(w))^2}\quad,
	\label{eq:Uth3}
\end{equation}
where $\operatorname{Re}$ denotes the real part.
The circulation $\Gamma_d$ in the disk plane is computed by integrating \cref{eq:IG}, an operation that we indicate by $\Gamma_d=\mathcal{I}(\Gamma_o)$.
Combining \cref{eq:Uth2,eq:Uth3} yields that the evolution of the vortex is described by the differential equation
\begin{equation}\label{eq:thiseq}
	\dfrac{dz_v}{dt} = \dfrac{1}{4\pi R} \, \mathcal{I} \left(2U(t) \dfrac{|g(R-iz_v)|^3}{\operatorname{Re}(g(R-iz_v))^2}\right) \mathcal{F}\left(\dfrac{z_v}{R}\right).
\end{equation}

The vortex circulation deriving from the integration of \cref{eq:thiseq} is presented in \cref{fig:fitE}.
The predicted circulation $\Gamma_o$ slightly overestimates the measured circulation during the transient stage.
This bias has multiple causes.
First, the estimation of the vortex velocity by \cref{eq:Uth} is only valid for steady vortices in unbounded fluids.
The presence of the plate and the transient nature of the flow can lead to overestimation of the vortex velocity.
Second, the selected mapping function $f$ is elegant but it is only one out of an infinite number of ways to map the disk.
The mapping induces a strong compression of the plane close to the tip of the disk (\cref{fig:mapmap}), which potentially leads to an inaccurate mapping of the vortex position in the open plane.
The lack of vorticity preservation by the mapping leads to a \SI{14}{\percent} decrease of the converged maximum circulation when mapped from the open plane to the disk plane.
Despite the reduced order of the model which does not consider the transient growth of the vortex, we obtain an excellent prediction of the maximum circulation of the vortex in the close wake of the cone.

\subsection{Discrete vortex method estimation of tail shedding}
\begin{figure}
\includegraphics{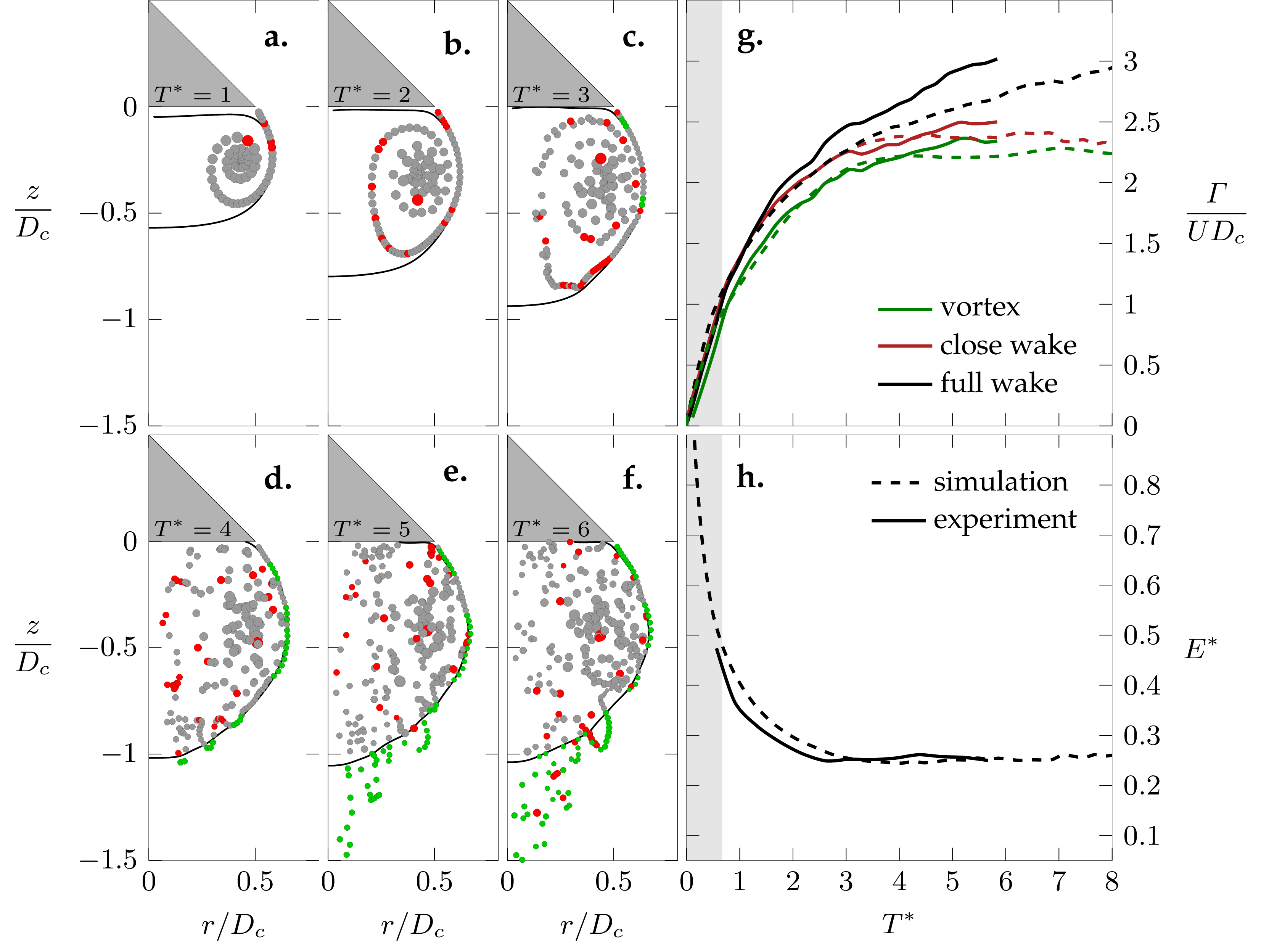}
\caption{\textbf{a-f}, Vortex development predicted by the discrete vortex method from $T^*=1$ to $T^*=6$.
The size of the point vortices increases with their circulation.
The black line indicates the vortex boundary, defined by the $\psi=0$ contour.
The colours indicate if a point vortex stays within the vortex boundary during the first eight convective times (grey), circumvents the vortex (green), or is released by tail shedding (red).
\textbf{g}, Temporal evolution of the simulated circulation (dashed lines) inside the vortex and in the full wake compared with experimental results (solid lines).
\textbf{h}, Temporal evolution of the non-dimensional energy of the main vortex ring.}
\label{fig:DVM_circ}
\end{figure}

The low order model based on conformal mapping presented in the previous section provides an accurate estimate of the maximum vortex circulation.
However, the model assumes that all produced vorticity ends up in the vortex and does not predict the vorticity lost by tail shedding.
To capture the effect of tail shedding, we apply here the discrete vortex model developed in \cref{sec:DVM}.
Simulations of the vortex growth in the wake of a cone with aperture $\alpha=\ang{45}$ are presented in \cref{fig:DVM_circ}.
As smoothing parameter in \cref{eq:smoothparam}, we used $\epsilon=D_c/10$ with a time step $dt=0.01D_c/U$.
The translation trajectory of the cone matches the experimental conditions for the cone with aperture $\alpha=\ang{45}$, disc diameter $D_c =\SI{6}{\centi\meter}$ and final velocity $U = \SI{0.5}{\meter\per\second}$.
The grey area indicates the acceleration phase of the cone.

The size of the discrete point vortices in \cref{fig:DVM_circ}a-f are a measure of their circulation.
We have also divided the vortices in three categories.
Point vortices that stay within the vortex boundary during the first eight convective times are presented in grey.
As the vortex boundary, we consider here the iso-contour of the stream function $\psi=0$.
Vortices that do not enter the vortex boundary at all and circumvents it are coloured in green.
Vortices that enter the vortex contour for some time but that are released via tail shedding are coloured in red.
In \cref{fig:DVM_circ}g, we represent the temporal evolution of the total circulation of all point vortices and the evolution of the circulation in the rectangular contours that cover the near wake and the full wake similar to what we did in \cref{fig:circ1}.

In the beginning of the translation, up to $T^*\approx 3$, all of the discrete point vortices released end up in the main vortex, delimited by the $\psi=0$ contour (\cref{fig:DVM_circ}a-c).
Some of them will be released again at a later stage via tail shedding (\cref{fig:DVM_circ}f).
A vortex is considered to be lost via tail-shedding if it has performed at least one rotation within the main vortex.
Most discrete vortices released after $T^*>3$ directly circumvent the vortex (\cref{fig:DVM_circ}d-f).

To calculate the temporal evolution of the total lost circulation, we calculate again the difference between the circulation within a rectangular contour overing the full wake and the circulation within the vortex boundary which is now defined by the iso-contour of the stream function $\psi=0$.
The evolution of the circulation bound by the different contours is presented in \cref{fig:DVM_circ}g.
The portion of the total lost circulation that is due to tail shedding is determined based on the classification of the point vortices.
At the end of the simulation, after eight convective times, vorticity lost by tail-shedding accounts for \SI{31}{\percent} of the total loss.
This percentage number is sensitive to the time step and the smoothing parameter.
When we divide or multiply the time step by eight, the fraction of the lost circulation due to tail shedding respectively reduces to \SI{20}{\percent} or increases to \SI{34}{\percent}.
When the smoothing parameter is reduced to $\epsilon=0$, the portion of circulation lost by tail-shedding increases beyond \SI{50}{\percent}.

Tail shedding originates from disturbances in the self similarity of the roll-up in the core of the vortex \citep{Krasny2002a}.
When we increase the time step for the simulation, we increase the spacing between consecutive point vortices.
If the distance further increases due to stretching of the vortex sheet near the core, it can become too large to accurately discretise the curvature of the shear layer and numerically amplifies the disturbance leading to tail-shedding.
A larger time step is thus expected to increase tail-shedding.
Decreasing the smoothing parameter destabilises the shear layer and has a similar effect as increasing the time step.

In our experiments, the Reynolds number based on the translation velocity and the diameter of the cone is higher than \num{10000} and the shear layer is expected to be turbulent (\cref{fig:setup}b).
Perturbations of the shear layer are observed early on in the roll-up process and quickly trigger the tail-shedding instability.
Simulating a perfectly smooth roll-up of the shear layer by refining the time step as proposed by \citet{Krasny2002a} is not representative of the presented experiment and the simulation parameters are tuned to approximate the experimental conditions.

The choice of the numerical parameters has a strong influence on the portion of the total circulation that is lost via tail shedding, but does not affect the amount of circulation that ends up in the vortex.
For a large range of parameter setting, our discrete vortex method consistently predicts a converged maximum circulation of $\Gamma/UD_c=2.4$ which is reached around $T^*=3$.
Our model shows that the feeding of the vortex ring is largely independent of the instability or irregularity of the shear layer.
This was also observed experimentally for vortices forming behind a flat disc with undulations along the edges \citep{Kaiser2020}.
The vortex circulation as well as the total circulation are underestimated by the model.
Our model predicts a linear increase of the total circulation for $T^*>3$, exactly as we observe experimentally.
However, the discrete vortex model predicts a growth rate of $\dot{\Gamma}=0.1U^2$ which is only half of the growth rate we measure experimentally.
In the experiment, the vortex and its circulation continue to grow and only the circulation non-dimensionalised by the diameter of the vortex converges to a limiting value (\cref{fig:circ1}a).
In absence of viscosity, the vortex reaches a limiting size and circulation and this affects the Kutta condition and the rate of increase of the total circulation.

Even though the discrete vortex method does not predict increase in the vortex diameter, we can still predict the trajectory of the vortex centre if we assume that the vortex self-induced velocity converges to the velocity of the cone.
In previous work \citep{DeGuyon2021}, we approximated the vortex velocity $U_0$ based on the circulation and the non-dimensional energy as
\begin{equation}
U_0=\dfrac{\Gamma}{2\pi R_v}\left(E^* \sqrt{\pi}+\dfrac{3}{4}\right)\quad.
\end{equation}
If we consider now the vortex as a single point vortex with a circulation that satisfies the Giesing-Maskell condition, we can predict its trajectory such that $U_0=U$ at all times provided that we know its non-dimensional energy $E^*$.
The non-dimensional energy of a vortex ring is linearly related to the relative standard deviation of the vorticity, it is a measure of the vorticity distribution that converges to a unique value \citep{DeGuyon2021}.

\begin{figure}
	\includegraphics{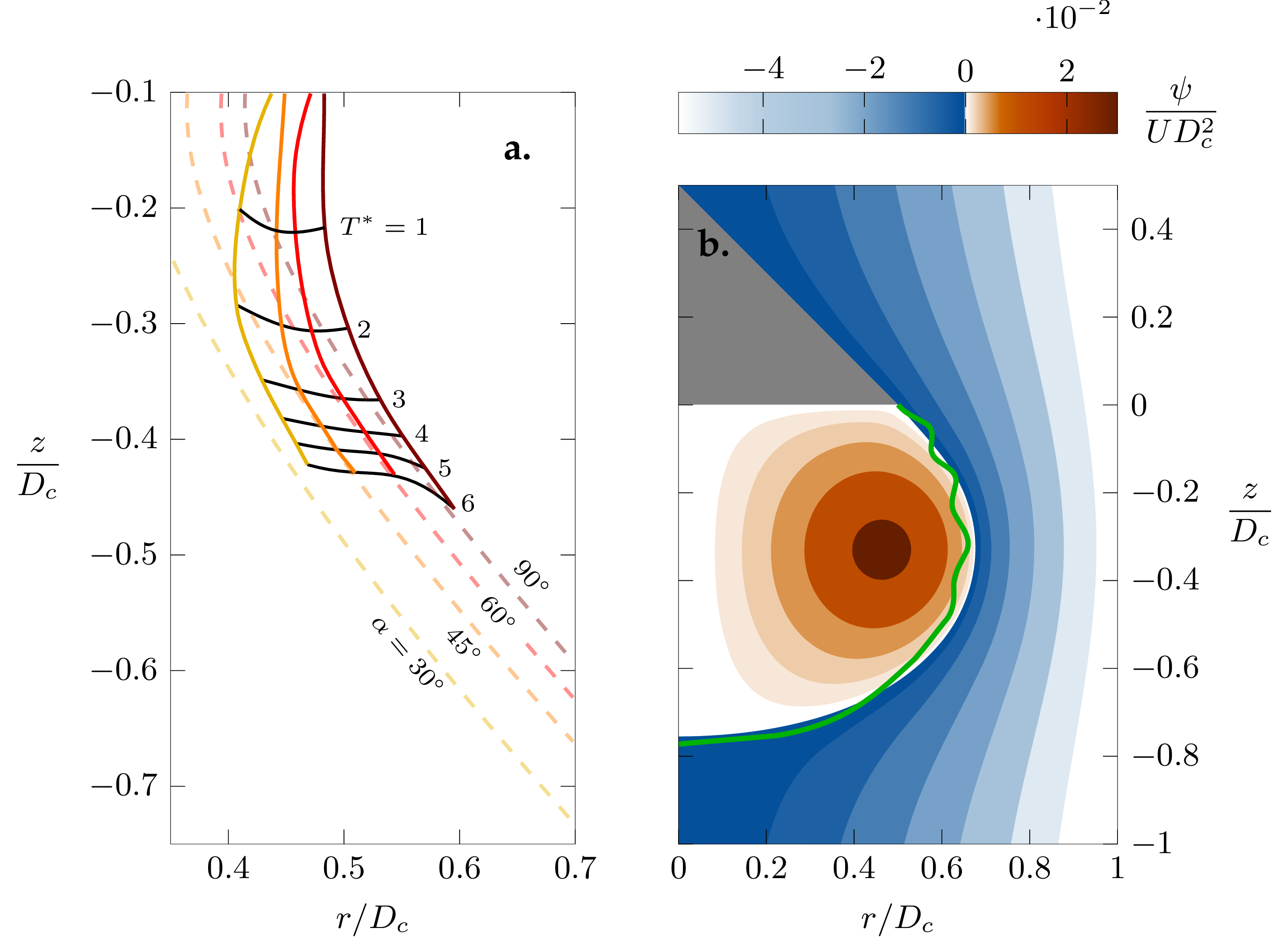}
\caption{\textbf{a}, Comparison between the measured trajectory of the vortex centre (solid lines) and its prediction using a single point vortex model and assuming $U_0=U$ (dashed lines).
\textbf{b}, Contours of the predicted stream function reconstructed using the position of the vortex at $T^*=3$.
The green line is the FTLE contour obtained from the experimental data.}
\label{fig:Psi_induced}
\end{figure}

The non-dimensional energy of the vortex behind the cone is obtained by calculating
\begin{equation}
E^*=\dfrac{E}{\sqrt{I \Gamma^3}} \qq{with} E=\pi \iint \psi \omega  \dd{r} \dd{z} \qq{and} I=\pi \iint \omega r^2 \dd{r} \dd{z}\quad.
\end{equation}
in the near wake of the cone.
The temporal evolution of the non-dimensional energy is presented in (\cref{fig:DVM_circ}h) for the experimental data and the discrete vortex model predictions.
The model accurately predicts the experimental evolution and both converge to a minimum limit value of $E^*=0.27$ for $T^*>3$.

Using this value for the non-dimensional energy of the vortex, we can estimate the trajectory of the vortex centre for various cone apertures by satisfying $U_0=U$ at every location.
The measured trajectories of the vortex centre converge asymptotically to the predicted paths for different cone apertures (\cref{fig:Psi_induced}a).
The vortex rings behind cones with smaller apertures initially contract and only grow larger than the cone base after four convective times when the predicted trajectory is linear.
Similar linear trajectories were observed for viscous vortex simulations behind a flat plate \citep{Johari2002}.
The simplified representation of the vortex ring as a single point vortex also provides a satisfying representation of the vortex dimension through comparison of the iso-contours of the predicted stream function with the experimentally determined pFTLE ridge (\cref{fig:Psi_induced}b).

\section{Conclusion}
Vortex rings occur in various shapes and sizes.
They are often created by ejecting a volume of fluid through a nozzle or they appear in the wake of accelerating bodies.
When fluid is ejected to form a vortex ring, its volume and velocity are known and this information is crucial in estimating the vortex' circulation, impulse, and energy by a slug flow model.
The vortex rings generated in this way move away from the nozzle that created them and they can be considered as unbound vortices.
When vortex rings form in the wake of an object, they enhance the fluid drag of the object and have a self-induced velocity towards the object that created them.
Low order models that aim at predicting the growth of these so-called drag vortices need to take the presence of the object into account and do not have a priori information on the volume and velocity of the fluid that will end up in the vortex.
In this paper, we addressed these challenges and presented two low order models to predict the growth of a vortex ring in the wake of translating disks or cones that are accelerated from rest.
The model results were compared with experimental time-resolved velocity field data.

The experimentally measured circulation in the wake of the cone starts to increases immediately when the cone is accelerated from rest and reaches a constant linear growth rate after three convective times for all cone geometries and Reynolds numbers considered.
The circulation normalised by the vortex diameter converges to a limiting maximum value around \num{2.3} and the vorticity distribution, quantified by the non-dimensional energy, reaches a minimum value around \num{0.3}.

The first model focusses on the prediction of the growth rate of circulation of an axisymmetric vortex ring behind a disk using conformal mapping.
The proposed mapping of the potential vector ensures the incompressibility of the flow and the non-penetration condition at the boundaries of the object, but does not preserve the irrotationality of axisymmetric flows.
A differential equation is derived for the vortex translational velocity based on its core radius and circulation, which is set to satisfy the Kutta condition at the tip of the disk.
The model accurately predicts the maximum vortex circulation but overestimates the circulation growth rate during the transient phase.
This is not unexpected as the model is based on Pullin's expression for the translational speed of steady vortex rings and due to large stretching of the mapping function near the tip of the disk.

The conformal mapping based model only predicts the temporal evolution of the total circulation which is assumed to end up entirely inside the vortex.
The experimental data revealed that the total circulation and the vortex circulation diverge after one to two convective times.
A portion of the vorticity generated at the walls of the object circumvents the vortex recirculation area, and around \SI{10}{\percent} of the vorticity in the far wake of the cone is lost via tail-shedding.
A second model in the form of an axisymmetric point vortex method was developed to account for the vorticity loss and to improve the predictions of the transient vortex growth.
The discrete vortex simulations predict well the temporal evolution of the circulation and non-dimensional energy of the vortex.
Due to its inviscid nature, the model does not capture the decreasing growth rate of the vortex after four convective times which is attributed to viscous effects.
The tail-shedding instability is well captured by the model.
The total amount of vorticity lost in the far wake of the cone is consistently predicted but the portion of it that is due to tail shedding is sensitive to the choice of the numerical parameters controlling the stability of the shear layer.
The increase in the time step and a decrease in the smoothing parameter numerically amplify the disturbance leading to tail-shedding.

 \subsection*{Declaration of interests.}
 The authors report no conflict of interest.

\bibliographystyle{jfm}
\bibliography{p2}
\end{document}